\documentclass[prl,twocolumn,a4paper,superscriptaddress,showpacs]{revtex4}

\usepackage{graphicx}
\usepackage{epsfig}
\usepackage{graphics,color}
\usepackage{amsmath}
\usepackage{amsfonts}
\usepackage{amssymb}
\usepackage{subfigure}

\begin{document}

\title{Electron-phonon interaction and full counting statistics
in molecular junctions}
\author{R. Avriller}
\email{avriller.remi@uam.es}
\author{A. Levy Yeyati}
\affiliation{Departamento de F\'{\i}sica Te\'orica de la Materia
Condensada C-V, Facultad de Ciencias, Universidad Aut\'onoma de Madrid,
E-28049 Madrid, Spain.}
\date{\today}
\begin{abstract}
The full counting statistics of a molecular level weakly interacting with a
local phonon mode is derived. We find an analytic formula that gives the behavior
of arbitrary irreducible moments of the distribution upon phonon
excitation. The underlying competition between quasi-elastic and inelastic processes
results in the formation of domains in parameter space characterized by a
given sign in the jump of the irreducible moments. In
the limit of perfect transmission, the corresponding distribution is distorted
from Gaussian statistics for electrons to Poissonian transfer of holes above the inelastic threshold.
\end{abstract}


\maketitle


It is now well established that the current noise generated by electronic
nanodevices contains valuable information on microscopic transport processes
not available from measurements of the current-voltage characteristics
\cite{Shot_Noise_Mesoscopic_Conductors}. A full characterization of the transport properties
of a device requires not only the knowledge of current-current correlations,
but rather the full counting statistics (FCS) has to be determined
\cite{Quantum_Noise_Mesoscopic_Physics}.
This amounts to determine the whole probability distribution $P_{t_0}(q)$ 
that a given charge $q$ is transmitted through the device during a certain
measurement time $t_0$. \\
Studies of FCS have been mainly restricted to non-interacting systems.
Notable examples of such studies are single channel conductors
\cite{Electron_Counting_Statistics_Coherent_States_Electric_Current}
or double quantum dot systems \cite{Counting_statistics_decoherence_coupled_quantum_dots}.
The case of FCS in the presence of electron-electron interactions in the coherent transport regime
 has been much less analyzed
 \cite{FCS_Single-Electron_Transistor_Nonequilibrium_Effects_Intermediate_Conductance}. In particular, the Kondo regime in quantum dots
has been addressed in Ref.\cite{FCS_Kondo_dot_Unitary_limit}. \\
On the other hand, molecular electronics
is becoming a field of intense research activity
\cite{Kondo_Resonances_Anomalous_Gate_Dependence_Electrical_Conductivity_Single_Molecule_Transistors}.  
In this case the coupling to vibrational modes plays an important role and 
provides an additional source for electronic correlations which may
affect the counting statistics \cite{FCS_Non_Ohmic_Transport_Molecules}. The case of atomic chains
suspended between metallic electrodes provides another test system
to analyze the effects of electron-phonon coupling in transport
properties \cite{Quantum_properties_atomic-sized_conductors}.
At low temperatures (quantum regime) the onset of phonon emission
processes is signaled by abrupt jumps in the system differential conductance
\cite{Universal_features_electron-phonon_interactions_atomic_wires, Unified_Description_Inelastic_Propensity_Rules_Electron_Transport_Nanoscale_Junctions}.   
When certain conditions are met, the behavior of the conductance jumps is entirely controlled by the 
transmission probability, evolving from a drop in conductance at high
transmission to an increase at low transmission. These predictions were
quantitatively confirmed in recent experiments \cite{Electron-Vibration_Interaction_Single-Molecule_Junctions}.
A natural question which arises concerns the behavior of noise
and, more generally, the FCS for energies corresponding
to the excitation of vibrational modes. Although some works have been
devoted to the analysis of noise in the presence of e-ph coupling
\cite{Theory_current_shot-noise_spectroscopy_single-molecular_QD_one_phonon_mode},
none of them tackled the problem of the determination of the FCS in the presence of
 e-ph interaction. \\
The aim of this work is to study how the FCS of a molecular junction is
modified by the coupling to a vibrational mode. On the basis of a simple
model, we derive a compact analytical expression encoding the FCS for the
experimentally relevant regime of weak e-ph interactions and strong coupling
to the leads, which corresponds to the conditions of
Ref.\cite{Electron-Vibration_Interaction_Single-Molecule_Junctions}. This expression allows to analyze the change of arbitrary
irreducible moments of the distribution $P_{t_0}(q)$ upon phonon excitation, as
well as giving a picture of the underlying interplay between quasi-elastic and
inelastic processes. \\
\noindent
The starting point of our derivation is the following model Hamiltonian 

\begin{flushleft}
\begin{eqnarray}  
&H& = \sum_{\mu} H_{\mu} + \epsilon_d\sum_{\sigma}d^{\dagger}_{\sigma}d_{\sigma} + \omega_{0} a^{\dagger}a + V + V_{e-ph} \\
&V& = \sum_{\mu, k, \sigma} t_{\mu d} \psi^{\dagger}_{\mu k \sigma}d_{\sigma}
+ \mbox{H.c} \mbox{ ; }   V_{e-ph} = \lambda (a +
a^{\dagger})\sum_{\sigma}d^{\dagger}_{\sigma}d_{\sigma} \nonumber
\end{eqnarray}
\end{flushleft}

\noindent
where a single molecular level of energy $\epsilon_d$ is coupled to the left (right)
electrode by a hopping element $t_{Ld} (t_{Rd})$, and interacts with a local
phonon mode of energy $\omega_{0}$ with e-ph coupling constant $\lambda$.
The indexes $(\mu, k, \sigma)$ label the state of the $\mu=L,R$ uncoupled
electrode, characterized by wave vector $k$ and spin $\sigma$. We further define the
cumulant generating function (CGF) $S(\chi) = -\sum_{n=1}^{+\infty}
\frac{(i\chi)^{n}}{n!}\big{<}q^{n}\big{>}_c$ as the functional generating the irreducible
moments of the distribution $\big{<}q^{n}\big{>}_c$. The connection of this definition to the former
Hamiltonian is given by \cite{Electron_Counting_Statistics_Coherent_States_Electric_Current}

\begin{eqnarray}  
e^{-S(\chi)} = \Big{<}T_c \exp \Big{\lbrace} -i \int_c V_{\chi(t)}(t) dt \Big{\rbrace} \Big{>}
\end{eqnarray}

\noindent
where $V_{\chi(t)}$ denotes $V$ with the substitution in the left
hopping element $t_{Ld}$ by $t_{Ld}e^{-i\chi(t)/2}$, and $T_c$ means
time ordering on the Keldysh contour going forward from time $0$ to time $t_0$
and backward from time $t_0$ to time $0$. The counting field $\chi(t)$ equals
to $\pm\chi$ on the forward (backward) branch of the Keldysh contour and accounts for a virtual
measurement of the charge being transmitted \cite{Quantum_Noise_Mesoscopic_Physics}.
As shown in Ref.\cite{FCS_Anderson_impurity_model}, it is convenient to work with
the generalized current $I(\chi)= s\frac{i}{t_0}\frac{\partial}{\partial \chi}
S(\chi)$, that can be expressed in terms of the Keldysh Green functions of the interacting molecular level
$G_{dd}^{\alpha\beta}(t,t')=-i \big{<}T_c d(t_{\alpha})
d^{\dagger}(t'_{\beta})\big{>}$ and of the uncoupled lead
$g_{LL}^{\alpha\beta}$ (with indexes $\alpha,\beta=\pm$) 

\begin{eqnarray}  
I(\chi) &=& \frac{s}{2\pi} W\Gamma_L \int d\omega \Big{\lbrace} e^{i\chi}
G_{dd}^{-+}(\omega) g_{LL}^{+-}(\omega) \nonumber \\
 &-& e^{-i\chi} G_{dd}^{+-}(\omega) g_{LL}^{-+}(\omega) \Big{\rbrace}
\end{eqnarray}

\noindent
In the former expression, $s=2$ stands for spin degeneracy and $\Gamma_L =
t_{Ld}^2/W$ is the coupling to the left contact expressed in units
of the inverse of density of states $W=1/\pi\rho_L$ (supposed to be
constant). 

\begin{figure}[!ht]
  \includegraphics[width=0.90\linewidth]{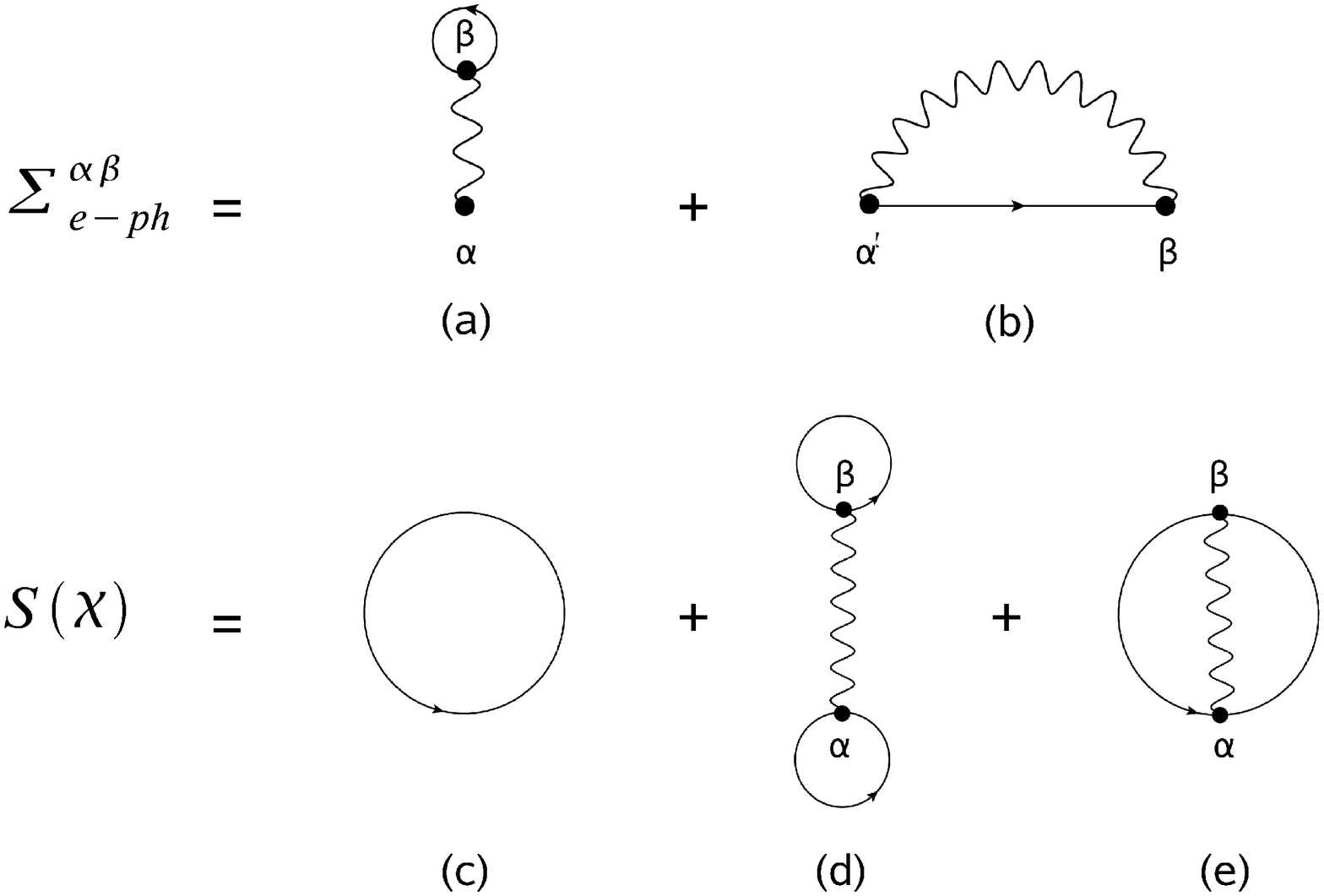} 
  \caption{Upper figure : Second order electron-phonon self energy in
    Keldysh space $(\alpha, \beta=\pm)$. (a) Hartree term. (b) Exchange term.
    Lower figure : corresponding bubble expansion of
    the CGF $S(\chi)$. (c) Unperturbed term. (d) Hartree term. (e) Exchange term.}
  \label{e-ph_self_energy_bubbles}
\end{figure}

\noindent
At second order in perturbation due to e-ph interaction, the dot
Green function can be written
$\hat{G}_{dd} \approx \hat{G}_{dd}^{(0)}+\hat{G}_{dd}^{(0)}\hat{\Sigma}_{dd}^{e-ph}\hat{G}_{dd}^{(0)}$,
and the problem of finding the CGF is thus equivalent to the one of computing e-ph
self-energy in Keldysh space $\hat{\Sigma}_{dd}^{e-ph}$ in the presence of the
counting field. We retain two
diagrams for the former. The first one is a Hartree like diagram
(Fig.\ref{e-ph_self_energy_bubbles}(a)) which is diagonal in Keldysh space,
frequency independent, of order $\lambda^2/\omega_0$, and
does not exhibit any jump at the inelastic threshold
($V=\omega_0$). The
second diagram is the exchange diagram (Fig.\ref{e-ph_self_energy_bubbles}(b))
that is responsible for the behaviour of transport properties at the inelastic
threshold. The corresponding bubble expansion of $S(\chi)$ is shown on
Fig.\ref{e-ph_self_energy_bubbles}(c) for the unperturbed CGF,
Fig.\ref{e-ph_self_energy_bubbles}(d) for the Hartree term and
Fig.\ref{e-ph_self_energy_bubbles}(e) for the exchange term.
Taking into account the Keldysh indexes, one obtains a natural decomposition
of $I(\chi)$ as $I_{0}(\chi) + I_{in}(\chi) + I_{el}(\chi)$, where
$I_{in}(\chi)$ is an inelastic contribution which arises from
the non diagonal elements of the e-ph self-energy ($\Sigma_{dd}^{+-}$ and
$\Sigma_{dd}^{-+}$), and $I_{el}(\chi)$ from the diagonal ones ($\Sigma_{dd}^{++}$ and
$\Sigma_{dd}^{--}$) \cite{EPAPS}. This decomposition is equivalent to
the one in Ref.\cite{Direct_calculation_tunnelling_current_Electron-phonon_interaction_effectsInteraction_Effects}. \\
\noindent
The unperturbed current $I_{0}(\chi)$ corresponds to resonant tunneling across
the molecular junction in absence of e-ph interaction, and is given by 

\begin{eqnarray}  
I_{0}(\chi) &=&  \frac{s}{2\pi} \int
\frac{d\omega}{\Delta_{\chi}}\Big{\lbrace}e^{i\chi}f_L(1-f_R)-
e^{-i\chi}f_R(1-f_L)\Big{\rbrace} \\
\Delta_{\chi}(\omega) &=& \frac{1}{T} + (e^{i\chi}-1)f_L (1-f_R)+(e^{-i\chi}-1)f_R (1-f_L) \nonumber
\end{eqnarray}

\noindent
where $f_{L(R)}$ is the Fermi distribution of the left (right) lead,
$T(\omega) = 4\Gamma_L\Gamma_R/\lbrack \Gamma^2+(\omega-\epsilon_d)^2\rbrack$ the zero bias transmission
coefficient and $\Gamma=\Gamma_L+\Gamma_R$ the total coupling to the
leads. The corresponding CGF coincides with
the one derived by Levitov \textit{et al.}
\cite{Electron_Counting_Statistics_Coherent_States_Electric_Current}.
Effects of e-ph interaction are included in the two remaining terms. The
inelastic contribution $I_{in}(\chi)$ can be written as 

\begin{widetext}
\begin{eqnarray}
I_{in}(\chi) &=& -\frac{s}{2\pi}\frac{2i}{\Gamma^2 T(\epsilon_d)} \int
\frac{d\omega}{\Delta_{\chi}} \Big{\lbrace} \Gamma_L \Big{\lbrack}
e^{i\chi}f_L\Sigma_{dd}^{-+} + e^{-i\chi}(1-f_L)\Sigma_{dd}^{+-} \Big{\rbrack}
\nonumber \\
&+& \frac{i}{\Delta_{\chi}} \frac{\partial \Delta_{\chi}}{\partial \chi} 
\Big{\lbrack} ( e^{i\chi}\Gamma_Lf_L +
\Gamma_Rf_R )\Sigma_{dd}^{-+} - ( e^{-i\chi}\Gamma_L(1-f_L) +
\Gamma_R(1-f_R) )\Sigma_{dd}^{+-} \Big{\rbrack} \Big{\rbrace}
\end{eqnarray}
\end{widetext}

\noindent
This corresponds to tunneling processes with absorption (emission) of a phonon.
During such a process, the final energy of the scattered electrons increases (decreases) by an amount
$\omega_0$ and the mean number of phonons decreases (increases) by one
unit. The last term $I_{el}(\chi)$ accounts for elastic processes during which
the energy of the scattered electrons is conserved, and is given by

\begin{eqnarray}
I_{el}(\chi) &=& -\frac{s}{2\pi} \frac{i}{\Gamma^2 T(\epsilon_d)} \int
\frac{d\omega}{\Delta_{\chi}^2} \frac{\partial \Delta_{\chi}}{\partial \chi} \Big{\lbrace}
(\omega-\omega_d)\Big{\lbrack} \Sigma_{dd}^{++} - \Sigma_{dd}^{--}
\Big{\rbrack} \nonumber \\
&+& i\Big{\lbrack} \Gamma_L(2f_L-1) + \Gamma_R(2f_R-1)
\Big{\rbrack} \Big{\lbrack} \Sigma_{dd}^{++} + \Sigma_{dd}^{--}
\Big{\rbrack} \Big{\rbrace}
\end{eqnarray}

\noindent
The term involving $\Sigma_{dd}^{++} - \Sigma_{dd}^{--}$ corresponds to
a renormalization of the transmission factor that gives logarithmic
corrections, whereas the term involving $\Sigma_{dd}^{++} + \Sigma_{dd}^{--}$
corresponds to quasi-elastic tunneling with emission-reabsorption of a
phonon (hence the phonon population is unchanged) together with a virtual
leaking of the propagating electrons into the leads. \\
The former compact formulas can be used to implement a
numerical computation of the FCS \cite{EPAPS}. Of particular interest is the behavior of
$I(\chi)$ at inelastic threshold, which is encoded in the jump of the
generalized conductance $\Delta G(\chi)=\frac{\partial}{\partial V}
I(\chi)\big{|}_{\omega_0^+}-\frac{\partial}{\partial
  V}I(\chi)\big{|}_{\omega_0^-}$. At zero temperature (assuming phonon population $f_B=0$), we
find an analytical formula for $\Delta G(\chi)$

\begin{eqnarray}
\Delta G(\chi) &=& \Delta G_{in}(\chi) + \Delta G_{el}(\chi) \\
\Delta G_{in}(\chi) &=& \frac{s}{2\pi}\lambda_{e-ph}^2
\frac{e^{i\chi}}{T(\epsilon_d)}\Big{\lbrace} \frac{1}{\Delta_{\chi;+}\Delta_{\chi;-}}-e^{i\chi}L_1
\Big{\rbrace} \nonumber \\
\Delta G_{el}(\chi) &=& \frac{s}{2\pi}\lambda_{e-ph}^2
\frac{e^{i\chi}}{T(\epsilon_d)}\Big{\lbrace}\frac{\frac{\omega_0}{2}^2-\epsilon_d^2}{4\Gamma_L\Gamma_R}\big{\lbrack}
L_2+L_3-L_1 \big{\rbrack}\nonumber \\
&-& \frac{\alpha-1}{4\alpha}\big{\lbrack} (\alpha-1)L_1
+ (\alpha+1)(L_3-L_2) \big{\rbrack} \Big{\rbrace} \nonumber
\end{eqnarray}

\noindent
where we have introduced the dimensionless e-ph coupling
$\lambda_{e-ph}^2=\lambda^2/\Gamma^2$ and the parameter $\alpha=\Gamma_L/\Gamma_R$ measuring the
asymmetry in the coupling to the leads. The following notation is adopted $\Delta_{\chi;\pm}\equiv \Delta_{\chi}(\pm
\frac{\omega_0}{2})$, $T_{\pm}\equiv T(\pm\frac{\omega_0}{2})$, $L_1=1/\Delta_{\chi;+}\Delta_{\chi;-}
\big{\lbrace} 1/\Delta_{\chi;+}+1/\Delta_{\chi;-}\big{\rbrace}$ and
$L_{2(3)}=T_{\mp}/\Delta_{\chi;\pm}^2$.

\begin{figure}
  \begin{center}
    \begin{tabular}{c}
      \includegraphics[width=0.807\linewidth]{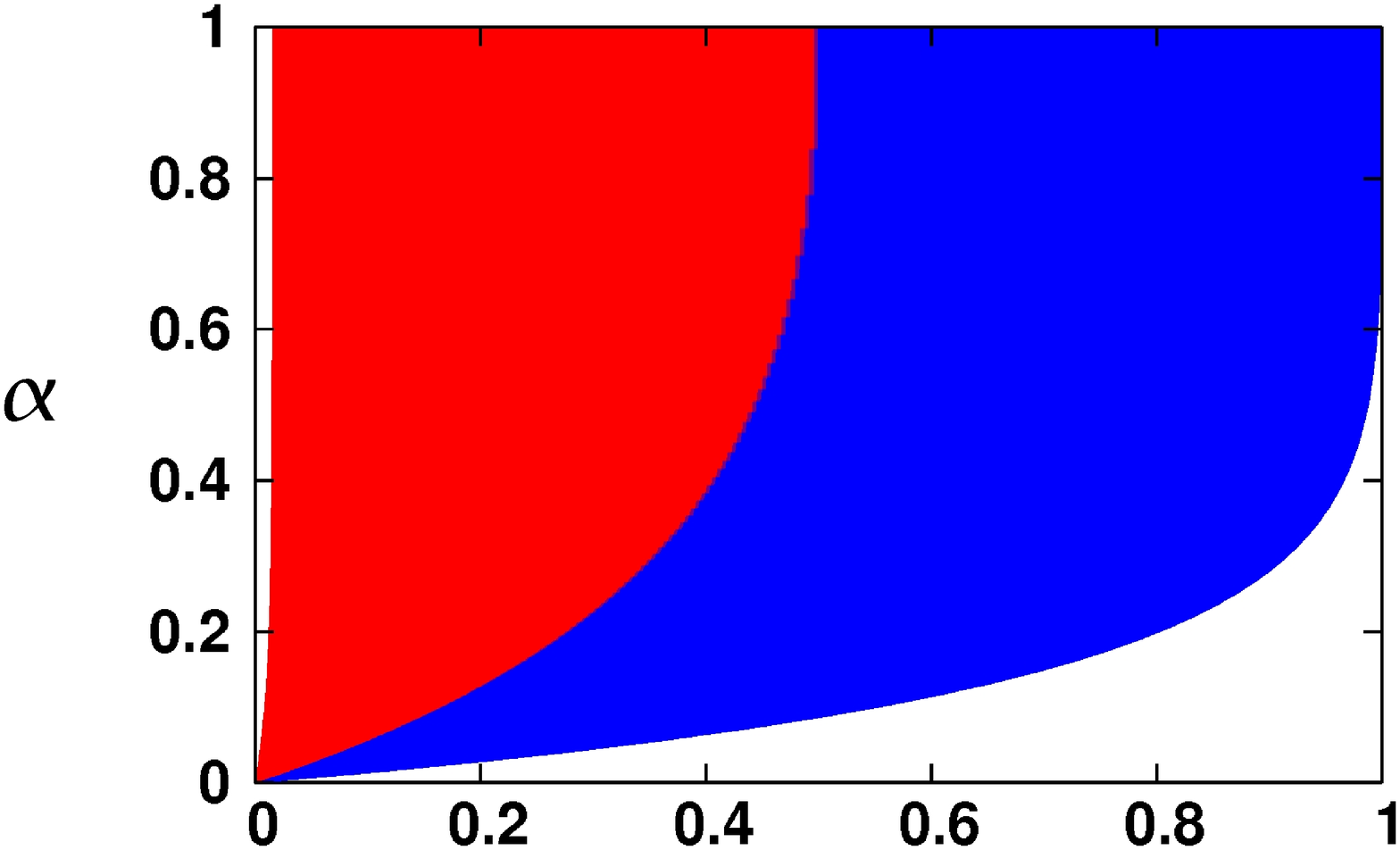} \\
      \includegraphics[width=0.800\linewidth]{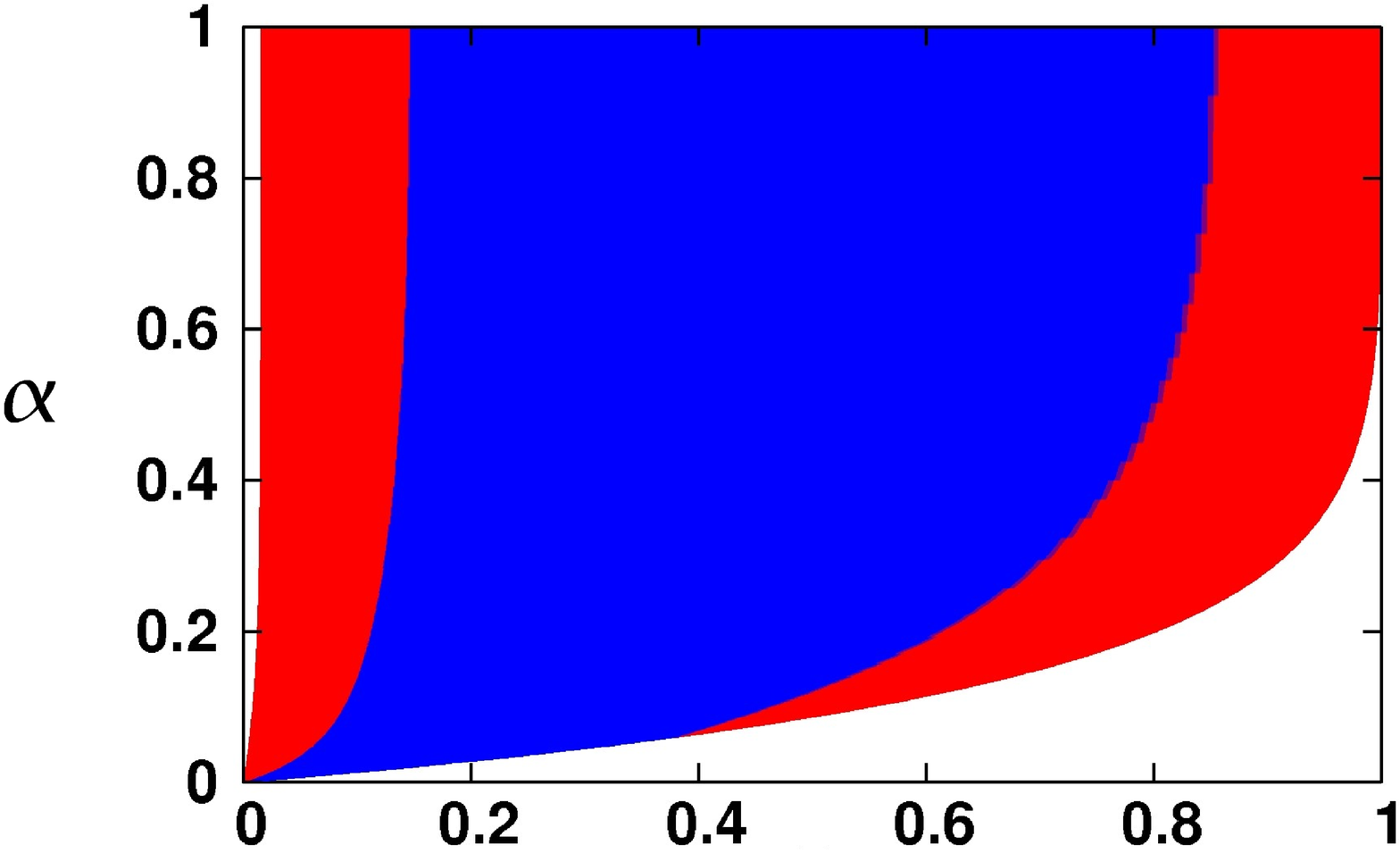} \\
      \includegraphics[width=0.807\linewidth]{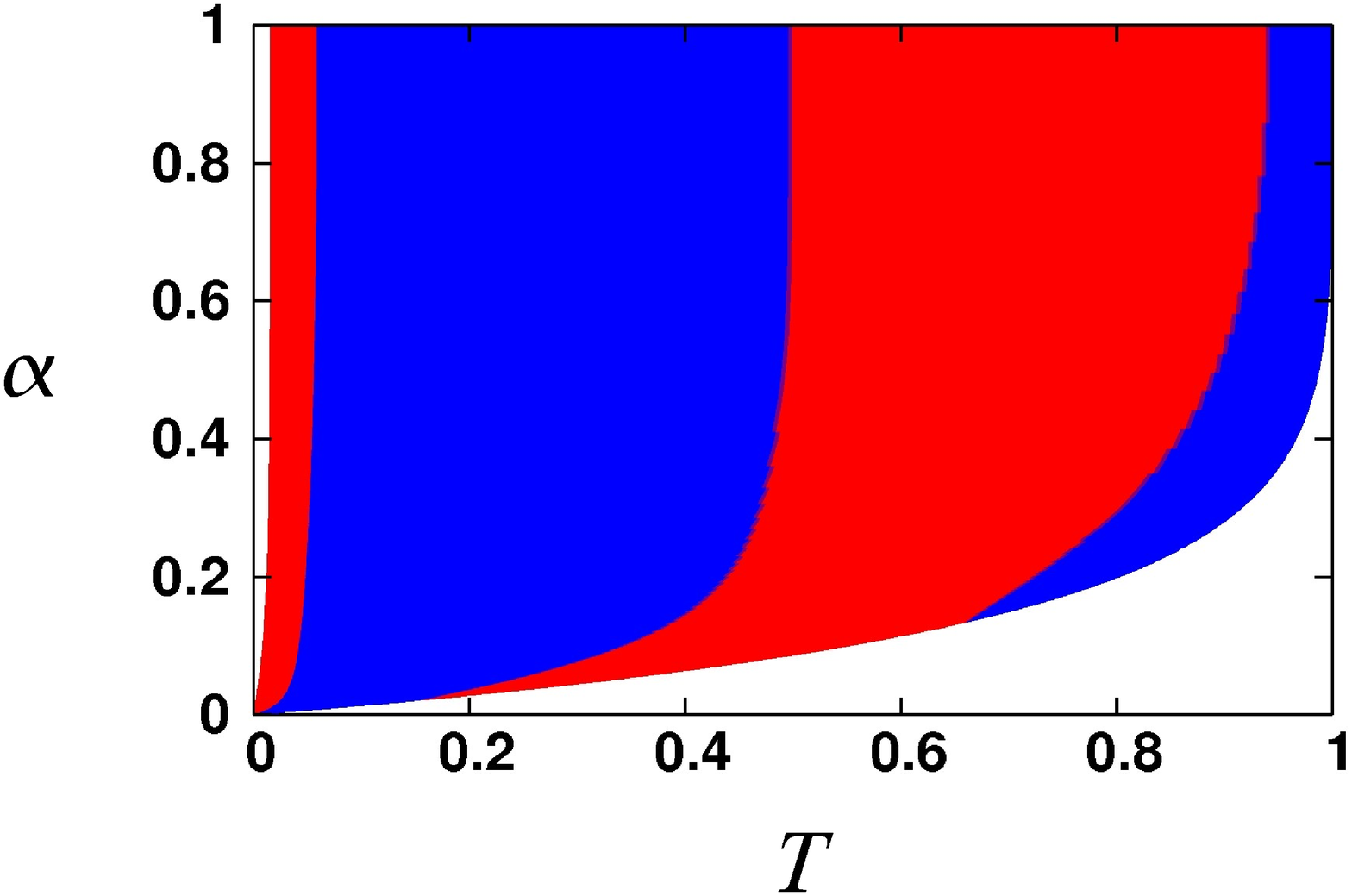} 
    \end{tabular}
    \caption{(Color online) Sign of the jump in the derivative with respect to voltage of the order $n$
      cumulant at phonon energy $\Delta F_n$.
      Parameter space is $\lbrace \alpha,T(0) \rbrace$, and phonon
      energy is $\omega_0=10^{-2}\Gamma$. From up to down :
      $\Delta F_1$ (jump in the conductance), $\Delta F_2$ (jump in the noise) and
      $\Delta F_3$ (jump in the third moment). The blue (red) color encodes a
      negative (positive) jump.}
    \label{Jump_Moment}
  \end{center}
\end{figure}

\noindent
This formula is the main result of the paper. We emphasize that it encodes the full energy
dependence of the model, \textit{i.e.} it is not restricted to wide band
approximation, and allows to explore the behavior 
of the derivative with respect to voltage of the order $n$ cumulant $F_n = \frac{1}{t_0 T(0) \lambda_{e-ph}^{2}} \frac{\partial}{\partial
V} \big{<}q^n\big{>}_c$, which exhibits a jump at phonon energy
given by   

\begin{eqnarray}
\Delta F_n = \frac{1}{i^{n-1} \lambda_{e-ph}^{2}} \frac{\partial^{n-1}}{\partial
\chi^{n-1}} \frac{\Delta G(\chi)}{T(0)}\Big{|}_{\chi=0}
\end{eqnarray}

\noindent
We show in Fig.\ref{Jump_Moment} the phase diagrams derived for the first
three $F_n$ factors, when exploring parameter space $\lbrace\alpha,T(0)\rbrace$
by modulating $\alpha=\Gamma_L/\Gamma_R$ and shifting the molecular level $\epsilon_d$. The first $F_1$
factor exhibits a jump at phonon energy $\Delta F_1$ (jump in the conductance), easily expressed as 

\begin{eqnarray}
\Delta F_{1;in} &=& \frac{s}{2\pi} \frac{T_{+}T_{-}}{T(0)T(\epsilon_d)}
\Big{\lbrace} 1 - (T_{+}+T_{-}) \Big{\rbrace} \\
\Delta F_{1;el} &=& -\frac{s}{2\pi} \frac{T_{+}T_{-}}{T(0)T(\epsilon_d)} \frac{\alpha-1}{2\alpha}
\Big{\lbrace} \alpha T_{-} - T_{+} \Big{\rbrace}
\end{eqnarray}

\noindent
The sign of $\Delta F_1$ (represented on upper panel of Fig.\ref{Jump_Moment}) has been
studied in Ref.\cite{Universal_features_electron-phonon_interactions_atomic_wires,
  Unified_Description_Inelastic_Propensity_Rules_Electron_Transport_Nanoscale_Junctions,Electron-Vibration_Interaction_Single-Molecule_Junctions}. We find two regions of the parameter
space corresponding to a negative jump (in blue) and a positive one (in red).
If $\omega_0 \ll \Gamma$ (which is the case for Fig.\ref{Jump_Moment}
where $\omega_0 = 10^{-2} \Gamma$), the contribution due to inelastic
processes $\Delta F_{1;in}$ is positive when $T(0) \leq 1/2$, due to the opening of an
 inelastic channel and negative when $T(0) \geq 1/2$, due to enhanced
 inelastic backscattering (the correction to this strong coupling behavior is of second order
 in $\omega_0/\Gamma$). On the other hand, the contribution of quasi-elastic
 processes to the jump $\Delta F_{1;el}$ (of first order in
 $\omega_0/\Gamma$) is always negative, due to elastic
 backscattering. Interestingly, $\Delta F_{1;el}$ is proportional to
$\alpha-1$ and exactly zero when the contact is symmetric ($\alpha =
1$). We emphasize that quasi-elastic processes contribute to the jump, because of  
virtual propagation into the electrodes during the
emission-reabsorption process, that is Pauli blocked when $V \leq
\omega_0$ (no available final scattering states). In the limit
$\alpha \rightarrow 0$, both quasi-elastic and inelastic processes are of the same
order of magnitude and fully compete. 
The case of the second factor $F_2$ (middle panel of Fig.\ref{Jump_Moment})
corresponds to the jump in the noise at phonon energy $\Delta F_2$

\begin{flushleft}
\begin{eqnarray}
&&\Delta F_{2;in} = \frac{s}{2\pi} \frac{T_{+}T_{-}}{T(0)T(\epsilon_d)}
\Big{\lbrace} 1 - 3(T_{+}+T_{-}) \\
&+& 2(T_{+}^2+T_{-}^2+T_{+}T_{-}) \Big{\rbrace}
\nonumber\\
&&\Delta F_{2;el} = \frac{s}{2\pi} \frac{T_{+}T_{-}}{T(0)T(\epsilon_d)}
\Big{\lbrace} \frac{T_{+} T_{-}}{2\Gamma_L\Gamma_R}
\big{\lbrack}\frac{\omega_{0}}{2}^2-\epsilon_d^2\big{\rbrack} \\ 
&-& \frac{\alpha-1}{2\alpha}\big{\lbrack} \alpha T_{-}(1-2T_{-}-T_{+}) -
T_{+}(1-2T_{+}-T_{-}) \big{\rbrack}\Big{\rbrace} \nonumber
\end{eqnarray}
\end{flushleft}

\noindent
We find three regions of parameter space.
For $\omega_0 \ll \Gamma$, and in the limiting case of a symmetric junction ($\alpha=1$), the
resulting total jump in the noise $\Delta F_{2}=\Delta F_{2;in}+\Delta F_{2;el}$ 
is positive when $T(0)\leq 1/2-1/2\sqrt{2}$ or $T(0)\geq 1/2+1/2\sqrt{2}$ and negative otherwise.
This change of sign can be understood by the following qualitative
arguments. In the regime where $T(0)$ goes to $0$ or $1$, shot noise goes to
zero due to Pauli principle, and activated e-ph interactions open
an inelastic channel (positive contribution to noise). In the intermediate regime where
$T(0) \approx \frac{1}{2}$, shot noise is maximal and activated
e-ph interactions result in a negative contribution to
noise. The same type of diagram is shown for the jump in the third moment $\Delta F_{3}$
(jump in the skewness) on lower panel of Fig.\ref{Jump_Moment}, where the competition between quasi-elastic and inelastic
processes results in the partition of parameter space in four regions. \\
\noindent
The behavior of an arbitrary $\Delta F_{n}$ and the whole FCS can be determined in the limit $T \rightarrow 1$,
where we obtain the following analytic approximation of the CGF \footnote{In
  the following equation, we have assumed $V>0$. The case $V<0$ is obtained
  by exchanging left and right indexes, and changing the sign of the counting
  field.} 

\begin{eqnarray}
S(\chi) \approx -i\overline{q}_0 \chi - \overline{q}_1 (e^{-i\chi}-1)
\end{eqnarray}

\noindent
where $\overline{q}_0 = t_0 V/2\pi$ and $\overline{q}_1 = t_0/2\pi
\lambda_{e-ph}^2/T(\epsilon_d) (V-\omega_0)\theta(V-\omega_0)$. Below the inelastic
threshold ($V < \omega_0$), the distribution
$P_{t_0}(q)$ is Gaussian (a delta peak at zero temperature) due to perfect transfer of mean charge
$\overline{q}_0$, whereas above that threshold ($V \geq \omega_0$),
$P_{t_0}(q)$ is distorted to a Poisson distribution for holes due to the activation of
sponteneous phonon emission (rare event for weak e-ph coupling). \\              
\indent
In conclusion, we have derived a compact formula for computing the FCS of a
molecular level weakly interacting with a local phonon mode. The competition between quasi-elastic and inelastic
processes results in the partition of the parameter space
$\lbrace\alpha,T(0)\rbrace$ into $n+1$ domains characterized by a given sign
in the jump of the generalized cumulant of order $n$ at phonon energy $\Delta F_n$. In the limit of perfect transmission, $P_{t_0}(q)$ evolves
from a Gaussian distribution for electrons to Poissonian distribution for holes, under activation of e-ph
interaction. Of immediate experimental interest is the change of sign in the jump of noise. For
temperatures in the range $T=4-17\mbox{ K}$, and typical energy of the 
phonon mode $\omega_0 \approx 50\mbox{ meV}$, the ratio $T/\omega_0\approx
10^{-2}-4.10^{-2}$ is very small and the jump is not smeared by thermal
effects. The amplitude of the jump being of order a few percent, we expect that the change of sign could be tested experimentally along the lines of
Ref.\cite{Electron-Vibration_Interaction_Single-Molecule_Junctions}. \\
The author aknowledge useful discussions with J.M. van Ruitenbeek, S. Bergeret and
J.C. Cuevas. Financial support from the Spanish MICINN under contract
NAN2007-29366-E is acknowledged. After submission, we became
aware of two closely related preprints \cite{Recent_References_FCS}.

\end{document}